\documentclass[aps,twocolumn,prl]{revtex4-1}

\usepackage{amsmath}
\usepackage{graphicx}
\usepackage{dcolumn}
\usepackage{bm}
\usepackage{nicefrac}

\def\apj{Astrophys. J.}
\def\apjss{Astrophys. J. Suppl. Ser.}

\def\epl{Europhys. Lett.}

\def\gca{Geochim. Cosmochim. Acta}

\def\lnp{Lect. Notes Phys.}

\def\mnras{Mon. Not. R. Astron. Soc.}

\def\pra{Phys. Rev. A}

\def\prc{Phys. Rev. C}
\def\prd{Phys. Rev. D}

\def\prl{Phys. Rev. Lett.}

\def\sci{Science}

\newcommand{\E}[1]{\ensuremath{\times 10^{#1}}}
\newcommand{\kmsec}{km\,s$^{-1}$}
\newcommand{\Cite}[1]{\mbox{Ref.~\cite{#1}}}

\newcommand{\eref}[1]{(\ref{#1})}
\newcommand{\Eref}[1]{Eq.~(\ref{#1})}

\begin{document}

\title{Manifestations of a spatial variation of fundamental constants on atomic clocks, Oklo, meteorites, and cosmological phenomena}

\author{J. C. Berengut}
\author{V. V. Flambaum}
\affiliation{School of Physics, University of New South Wales, Sydney 2052, Australia}

\date{23 August 2010}

\begin{abstract}

The remarkable detection of a spatial variation in the fine-structure constant $\alpha$ from quasar absorption systems~\cite{webb10arxiv} must be independently confirmed by complementary searches. In this letter, we discuss how terrestrial measurements of time-variation of the fundamental constants in the laboratory, meteorite data, and analysis of the Oklo nuclear reactor can be used to corroborate the spatial variation seen by astronomers. Furthermore, we show that spatial variation of the fundamental constants may be observable as spatial anisotropy in the cosmic microwave background, the accelerated expansion (dark energy), and large-scale structure of the Universe.

\end{abstract}

\maketitle

\section{Introduction}
\label{sec:intro}

A very large study of quasar absorption systems has recently detected a spatial variation in the fine-structure constant, $\alpha = e^2/\hbar c$~\cite{webb10arxiv}. The general idea is to compare the wavelengths of atomic spectra measured in the laboratory with those seen in absorption systems at high redshift. Any variation in the value of $\alpha$ results in well-understood discrepancies between the two spectra. Systematics are controlled to a very high degree by using many atoms and ions~\cite{dzuba99prl,dzuba99pra}.

Previous studies of quasar absorption spectra had suggested that the fine-structure constant may have been smaller in the past~\cite{webb99prl,murphy01mnrasA,murphy03mnras,murphy04lnp}. However, these studies all used spectra taken at the Keck telescope in Hawaii, at a latitude of $20^\circ$~N. Similar studies using the Very Large Telescope in Chile (latitude $25^\circ$~S) at first showed a stringent null constraint~\cite{srianand04prl}. More careful analysis of the same sample suggested that the errors should be enlarged by a factor of six~\cite{murphy07prl,murphy08mnras}, and that a much larger, dedicated VLT survey be performed.

The most recent work, which makes use of both Keck and VLT data, shows a highly significant ($\sim 4\sigma$) spatial gradient in the value of $\alpha$~\cite{webb10arxiv}. That is, $\alpha$ was larger in the past in one direction and smaller in the past in the opposite direction. This dipole, which we will refer to as the ``Australian dipole'', has a declination of around $-60^\circ$. This explains why the previous studies suggested a time-varying $\alpha$ that was smaller in the past: they only used data from the Keck telescope, which sees mainly in the Northern Hemisphere. The new results are entirely consistent with the previous ones.

Discovery of a spatial variation in the fundamental constants of nature has massive implications for the ``fine-tuning'' problem. This is the question of why the constants of nature seem to be finely-tuned for life to exist. While the anthropic principle can be invoked to explain such tuning of the Universe, there remains the question of how it could come about. With the detection of spatial variation of constants we begin to have a natural explanation for fine-tuning: with many possibilities for combinations of constants all occurring within the (possibly infinite in extent) Universe, we simply appear in the part of the Universe that is consistent with our existence.

Extraordinary claims require extraordinary evidence, and the detection of a spatial gradient in $\alpha$ is no exception. In this letter we discuss how limits from atomic clock measurements, meteorite data, and the Oklo nuclear reactor may be interpreted in the light of the evidence for spatial variation. We show how these independent methods could be used to confirm or contradict the dipole model of $\alpha$-variation that the quasar data suggest. We also briefly discuss the possibility that a spatial variation may be observed via asymmetries in the expansion of the Universe, the cosmic microwave background, and the large-scale structure.

\section{Model}

The ``Australian dipole'' of $\alpha$-variation found by \cite{webb10arxiv} is
\begin{equation}
\label{eq:ausdipole}
	\frac{\delta\alpha}{\alpha_0} = (1.10\pm0.25)\E{-6}\, r \cos \psi
\end{equation}
where $\delta \alpha/\alpha_0 = (\alpha(\mathbf{r})-\alpha_0)/\alpha_0$ is the relative variation of $\alpha$ at a particular place $\mathbf{r}$ in the Universe (relative to Earth at $\mathbf{r}=0$). The function $r\cos\psi$ describes the geometry of the spatial variation: $\psi$ is the angle between the direction of the measurement and the axis of the Australian dipole, ($17.4\,(0.6)$~h, $-58\,(6)^\circ$) in equatorial coordinates. The distance function is the light-travel distance $r = ct$ measured in giga-lightyears. This is model dependent for large redshifts: we use the standard $\Lambda_\textrm{CDM}$ cosmology parametrized by WMAP5~\cite{hinshaw09apjss} to determine the light-travel time $t$. It is assumed here that $\delta\alpha/\alpha_0 = 0$ at zero redshift, which is supported by the data, however this assumption should be tested using the same absorption methods as are used at high redshift (e.g. by using absorbers within our own galaxy).

Our goal is to provide a simple interpretation of terrestrial measurements of variation of constants in terms of the spatial gradient in values of $\alpha$. A minimal hypothesis is to expect all fundamental constants to vary along the direction specified by the Australian dipole of~\Cite{webb10arxiv}. There are some good theoretical justifications for this postulation. For instance, the constants may vary because they are coupled to a (dimensionless) scalar field $\Phi$ which varies over space-time. For example the quintessence field $\Phi/c^2$ or a dimensionless dilaton field. In this case the axis of the dipole is the direction of its gradient $\nabla\Phi$, and a fundamental constant $X$ is coupled to its variation via
\begin{equation}
\frac{\delta X}{X_0} = k_X \delta \Phi
\end{equation}
where $k_X$ is a dimensionless coupling coefficient. Our dipole model now requires
$\delta \Phi (\mathbf{r}) \sim r\cos\psi$
but all constants will vary in the same direction (i.e. along the Australian dipole).

In this letter we will deal with the constants $\alpha=e^2/\hbar c$, the electron-to-proton mass ratio $\mu=m_e/m_p$, and the dimensionless mass ratio $X_q = m_q/\Lambda_\textrm{QCD}$ where $m_q$ is the light-current quark mass and $\Lambda_\textrm{QCD}$ is the position of the Landau pole in the logarithm of the running strong coupling constant, $\alpha_s \sim 1/\ln (\Lambda_\textrm{QCD}r/\hbar c)$. In the Standard Model the electron and quark masses are proportional to the vacuum expectation of the Higgs field, while the proton mass $m_p$ is proportional to $\Lambda_\textrm{QCD}$ (if we neglect the $\sim 10\%$ contribution of the quark masses). Relative variation of $X_q$ is then approximately equal to the relative variation of $m_q/m_p$ and $\mu = m_e/m_p$. We can relate the relative variation of different constants by equations like
\begin{equation}
\label{eq:k_mu_alpha}
	k_\mu = R^{\alpha}_{\mu}\, k_\alpha
\end{equation}
where the $R_X^{X'}$ can be determined from observations and compared with theories of spatial variation.

\section{Atomic clocks}
\label{sec:lab}

Laboratory-based atomic clocks provide measurements of the time-variation of $\alpha$ in the Earth frame. Since the Earth is moving with respect to the rest frame of the cosmic microwave background (CMB), we may expect $\alpha$ to vary as the Earth moves along the Australian dipole. The assumption here is that the dipole is stationary in the CMB rest-frame, which is to be expected if the variation is caused by a co-moving scalar field, for example. To start, we neglect the annual motion of the Earth around the Sun, which is averaged out over time. The velocity of the Sun in the CMB rest-frame is known to high accuracy from the CMB dipole itself, and is 369~\kmsec\ in the direction ($168^\circ$, $-7^\circ$)~\cite{hinshaw09apjss}. This is almost perpendicular to the direction of the Australian dipole, and is therefore sensitive to the exact angle, which has some uncertainty. We can expect the yearly variation in laboratory measurements to be
\begin{equation}
\label{eq:lab}
{\dot\alpha}/{\alpha}|_\textrm{lab} = 1.35\E{-18}\,\cos\psi~\text{yr}^{-1}
\end{equation}
where $\psi$ is the angle between the motion of the Sun and the dipole. The best fit value for the Australian dipole gives $\cos\psi \sim 0.07$, but this has an uncertainty $\sim 0.1$. This signal will be modulated by the annual motion of the Earth around the Sun; with the angle between the ecliptic plane and the Australian dipole taken to be $35^\circ$, this modulation is
\begin{equation}
\frac{\delta\alpha}{\alpha} = 1.4\E{-20}\,\cos \omega t
\end{equation}
where $\omega$ refers to the angular frequency of the yearly orbit.

\Eref{eq:lab} gives a useful benchmark for comparing laboratory measurements to the recent spatial $\alpha$-variation data. The current best limit on the rate of $\alpha$-variation in laboratory measurements is $\dot\alpha/\alpha = (-1.6\pm 2.3)\E{-17}~\text{yr}^{-1}$, obtained by comparison of Hg$^+$ and Al$^+$ clocks over the course of a year~\cite{rosenband08sci} using the proposal and calculations of \cite{dzuba99prl,dzuba99pra}. Therefore this limit will have to be improved by two orders of magnitude to compete with the astronomical spatial-variation data. Fortunately, atomic clocks are improving rapidly, and there are several new schemes that could allow measurement at this level of precision, including the $^{229}$Th nuclear clock proposal~\cite{peik03epl} that has a hugely enhanced sensitivity to $\alpha$-variation~\cite{flambaum06prl}.

To avoid misunderstanding, we note that the astronomical observations do not exclude temporal variation of $\alpha$ below the rate $10^{-16}$ per year. That is, when considering temporal variations, the laboratory observations are already competitive at their present level of accuracy.

\section{Oklo}
\label{sec:oklo}

We can get an estimate for the kind of variation expected at the Oklo natural nuclear reactor by seeing how far our local galaxy cluster has moved relative to the rest frame of the cosmic microwave background. Currently the center of our galaxy is moving at $552$~\kmsec\ in the direction ($266.5^\circ$, $29.1^\circ$) in galactic coordinates or ($158^\circ$, $-24^\circ$) in equatorial coordinates~\cite{kogut93apj}. Extrapolating this velocity over the $\sim 1.8$ billion years since the Oklo reactor was running gives a total movement of $3.3\E{6}$~light years. Of course, this does not account for any particular accelerations of the Milky Way galaxy with respect to the local group. Indeed, the local group as a whole is moving at around 630~\kmsec\ in roughly the same direction: this gives an indication of the expected deviation from constant velocity. 

The galactic center moves at an angle $\cos\psi \approx 0.24$ to the direction of the Australian dipole, so it has moved a distance of $0.78\E{6}$~light years along the dipole direction since the time of Oklo. Therefore, if the quasar results are correct, we expect that $\alpha$ will have been smaller on Earth in the past. To achieve accuracy from Oklo comparable to the quasar data will require determination of $\alpha$-variation at the time of Oklo at the level $\delta\alpha/\alpha \sim 1.1\E{-6} \times 0.78\E{-3} \sim 10^{-9}$. The most sensitive Oklo limits on variation are based on measurement of the position of a very low energy resonance ($E_r=0.1$~eV) in neutron capture by the $^{149}$Sm nucleus. Using the result~\cite{flambaum09prc}
\begin{equation}
\label{eq:deltaEr}
	\frac{\delta E_r}{\text{MeV}} = 10\frac{\delta X_q}{X_q} - \frac{\delta\alpha}{\alpha}
\end{equation}
we expect that $E_r$ would have been different at the time the Oklo reactor was running by
\begin{equation}
  \delta E_r|_\textrm{Oklo} \approx 1~\textrm{meV}\ - R^{\alpha}_q\, 10~\textrm{meV}
\end{equation}
where $R^{\alpha}_q$ is defined by $k_q = R^{\alpha}_q k_{\alpha}$ as in \Eref{eq:k_mu_alpha}. The Sm resonance has an order of magnitude more sensitivity to variation in the dimensionless light-quark mass $X_q$ than it does to $\alpha$. The current best $2\sigma$ limits from Oklo data are $-12 < \delta E_r < 26$~meV~\cite{gould06prc} and $-73 < \delta E_r < 62$~meV~\cite{petrov06prc}.

\section{Meteorites}
\label{sec:meteor}

Meteorites can be used to determine average rates of decay of long-lived isotopes over the last 4.6~gigayears, which can be compared with laboratory rates to test for changes in fundamental constants~\cite{dyson67prl,fujii03prl}. The energy difference in the $\beta$-decay of $^{187}$Re to $^{187}$Os is very small, $\omega = 2.66$~keV. It is the result of cancellation between the relatively large Coulomb and asymmetry energy differences (which are $\sim 20$~MeV~\cite{dyson67prl,olive02prd}) of the two nuclei, and so it is very sensitive to possible variation of constants.

We define our dimensionless observable by $y = \lambda_{Re}/\lambda_{U}$: the ratio of the decay rate of the $^{187}$Re isotope to that of a relatively insensitive isotope, such as Uranium. The relative variation of $y$ is $\delta y/y = \delta \lambda_{Re}/\lambda_{Re} - \delta \lambda_{U}/\lambda_{U}$. Since $\lambda_{Re} \ll \lambda_{U}$, the Uranium isotope serves to calibrate the time that the meteorite was formed. Therefore, taking $\lambda_U$ as fixed, we can express the sensitivity of $\lambda_{Re}$ to variation of $\alpha$ as $\delta \lambda/\lambda = K_\lambda^\alpha\, \delta\alpha/\alpha$, with $K_\lambda^\alpha \simeq 2\E{4}$~\cite{olive02prd,dent08prd}.

The decay rate determined from the Re-Os isochrons~\cite{smoliar96sci} is an average value over the time since the meteorites were formed,
\begin{equation}
\label{eq:lambdabar}
\bar\lambda = \frac{1}{t_0 - t_1} \int_{t_1}^{t_0} \lambda (t)\, dt\ ,
\end{equation}
where $t_1$ and $t_0$ are the meteorite time and present time, respectively. As the solar system moves through the CMB rest-frame along the Australian dipole, we expect the fine-structure constant to vary as \eref{eq:ausdipole} with $r\cos\psi = vt = 4.3\E{-4}\,t$ where $t$ and $r$ are measured in Gyrs. As in the Oklo case, this movement is that of the galaxy as a whole. Then
\begin{equation}
\frac{\delta \lambda (t)}{\lambda} = 4.8\E{-10}\,K_\lambda^\alpha \, t
\end{equation}
and integrating \eref{eq:lambdabar} with $t_0 - t_1 = 4.6$~Gyr gives
\begin{equation}
\bar\lambda = \lambda_0 \left( 1 - 1.1\E{-9}\,K_\lambda^\alpha \right)
\end{equation}
where $\lambda_0 = \lambda (t_0)$. Therefore the relative difference between the decay rates measured from meteorites and that measured in the lab is expected to be at the level 
\begin{equation}
\frac{\bar\lambda - \lambda_0}{\lambda_0} = -2.2\E{-5}\ .
\end{equation}
Unfortunately, this is far below the current accuracy, \mbox{$\sim 10^{-2}$}, which comes from $\bar\lambda = 1.666\,(9)\E{-11}~\text{yr}^{-1}$~\cite{smoliar96sci} and $\lambda_0 = 1.68\,(5)\E{-11}~\text{yr}^{-1}$~\cite{galeazzi00prc} (see also~\cite{lindner89gca}). Both of these measurements will need to improve significantly before the expected range of variation is reached.

\section{Cosmological implications}

Variations of the fundamental constants will lead to changes in the masses and binding energies of elementary particles, including leptons and baryons as well as dark matter particles. If one assumes cosmological conservation laws apply in a flat Universe, then
\[
\Omega_\textsl{baryons} + \Omega_\textsl{dark\ matter} + \Omega_\textsl{radiation} + \Omega_\textsl{dark energy} = 1\ ,
\]
and there must be an exchange between the light and dark matter energy densities and the ``dark energy''. Therefore a cosmological anisotropy in the fundamental constants will result in an anisotropic energy exchange between these contributions to the total energy density of the Universe.

The contribution of baryonic matter to gravitational processes will be approximately proportional to the dimensionless constant $G m_p^2/\hbar c \sim G\Lambda_\textrm{QCD}^2/\hbar c$. In the absence of a confirmed theory for dark matter, we may write a similar expression for the dark matter contribution, $G m_\textrm{DM}^2/\hbar c$, and a cross-term $G m_p m_\textrm{DM}/\hbar c$, which depend on an unknown mass $m_\textrm{DM}$. The effect of variation of these contributions to the energy density of the Universe may be related to varying $G$ theories.

We see that the observed $\alpha$-variation is related to mass density and the cosmological ``constant'', and hence the accelerated expansion of the Universe. Therefore it may be possible to see a spatial anisotropy in the redshift-luminoscity relationships of SnIa supernovae data. Additionally the temperature anisotropy of the cosmic microwave background and large-scale structure formation may have observable spatial anisotropies.

\section{Conclusion}

We have shown how laboratory experiments of time-variation of fundamental constants and limits from the Oklo nuclear reactor and meteorites can be compared to the spatial variations seen by astronomers. This interpretation is essential if one wishes to independently corroborate the spatial variation seen by astronomers. None of the current terrestrial limits contradict the reported observation of a spatial gradient in $\alpha$.

We have also tested other astrophysical systems where one could find spatial variation of fundamental constants: full results will be deferred to a later paper. Briefly, the existing $H_2$ data from quasar absorption spectra shows hints that there may be a dipole in $\mu$-variation with an axis corresponding to the $\alpha$-variation of \Cite{webb10arxiv}. On the other hand the variation of $x=\alpha^2\mu g_p$, inferred from 21-cm data, has a best-fit dipole whose axis does not correspond to that of the Australian dipole, although in this case systematics heavily dominate. It is also possible to infer spatial variation of fundamental constants during big bang nucleosynthesis from high-redshift measurements of primordial abundances. The existing deuterium data does not support the dipole interpretation statistically, but the preferred axis is aligned with the Australian dipole. There is a strong impetus now to perform measurements of relative primordial abundance at high redshifts of as many elements as possible in as many different spatial directions as possible.

Finally we note that it may be possible to observe a spatial dipole in other cosmological systems. For example, $\alpha$-variation may be seen in the CMB anisotropy if a high-enough sensitivity can be reached. Although the results of \cite{webb10arxiv} (interpreted as strictly spatial variation) suggest that accuracy at the level $10^{-6}$ will be required, if the hints from big bang nucleosynthesis turn out to be real then there is an additional redshift (time)-dependence that could increase the variation at the time of the CMB substantially. Another possibility is that since the observed $\alpha$-variation is related to variation in the mass density and cosmological constant, and hence the accelerated expansion of the Universe, it may be possible to see a dipole in the redshift-luminoscity relationships of SnIa supernovae data.

\acknowledgments
We thank G. t'Hooft, J. A. King, J. K. Webb, and M. T. Murphy for useful discussions. This work is partly supported by the Australian Research Council.

\bibliography{references}

\end{document}